\documentclass[prb, preprint, superscriptaddress]{revtex4-1}

\usepackage{amsmath}  
\usepackage{amsfonts} 
\usepackage{graphicx} 
\usepackage{subcaption}
\usepackage{bm}
\usepackage{placeins}
\usepackage{float}
\usepackage{setspace}
\usepackage{svg}
\usepackage[arrowdel]{physics}
\usepackage{makecell}
\usepackage[style=mmddyyyy]{datetime2}
\usepackage{booktabs, multirow} % for borders and merged ranges
\usepackage{soul}% for underlines
\usepackage{xcolor,colortbl} % for cell colors
\usepackage{changepage,threeparttable} % for wide tables
\usepackage{tablefootnote}
\usepackage{ifthen}
\newcommand{\nomUnc}[4]{$#1#2#3\ \mathrm{#4}$}
\newcommand{\chargeMassOne}{$\gamma_{1} = -1.57(5) \times 10^{-3}\ \mathrm{C/kg}$} 
\newcommand{\stdCmone}{$\sigma(\gamma_{1}) = 0.04 \times 10^{-3}\ \mathrm{C}/\mathrm{kg}$}
\newcommand{\chargeMassTwo}{$\gamma_{2} = -1.63(4) \times 10^{-3} \ \mathrm{C/kg}$} 
\newcommand{\stdCmTwo}{$\sigma (\gamma_{2}) = 0.8 \times 10^{-3}\ \mathrm{C}/\mathrm{kg}$}
\newcommand{\chargeMassmean}{$\mu(\gamma) = -2.1 \times 10^{-3}\ \mathrm{C/kg}$}

\newcommand{\stdcm}{$\sigma (\gamma) = 0.08  \times 10^{-3}$}

\newcommand{\meanACnull}{$4.64(34)\ \mathrm{mm}$}
\newcommand{\modelACnull}{$4.75(17)\ \mathrm{mm}$}

\newcommand\segSegspace{$1.0(1)\ \mathrm{mm}$} %Segmented-segmented electrode spacing
\newcommand\segACspace{$2.0(1)\ \mathrm{mm}$} %Segmented-AC electrode spacing 
\newcommand\ACGNDspace{$2.0(1)\ \mathrm{mm}$} %AC-GND electrode spacing
\newcommand{\highV}{\nomUnc{-4.95}{(4)}{\times 10^{2}}{V}}
\newcommand{\lowV}{\nomUnc{-2.59}{(5)}{\times 10^{2}}{V}}
\newcommand{\offV}{\nomUnc{-1}{(5)}{\times 10^{-2}}{V}}
\newcommand{\relayT}{\nomUnc{4.2}{(5)}{}{ms}}
\newcommand{\centVmin}{\nomUnc{-39}{(1)}{}{V}}
\newcommand{\centVmax}{\nomUnc{-300}{(2)}{}{V}}
\newcommand{\dimCentw}{\nomUnc{139.6}{(1)}{}{mm}}
\newcommand{\dimCentl}{\nomUnc{3.2}{(1)}{}{mm}}
\newcommand{\dimACw}{\nomUnc{139.6}{(1)}{}{mm}}
\newcommand{\dimACl}{\nomUnc{4.2}{(1)}{}{mm}}
\newcommand{\dimSegw}{\nomUnc{18.9}{(1)}{}{mm}}
\newcommand{\dimSegl}{\nomUnc{15.5}{(1)}{}{mm}}
\newcommand{\rChione}{$\hat{\chi}^{2}_{1} = 3.22$}
\newcommand{\rChitwo}{$\hat{\chi}^{2}_{2}  = 3.48$}
\newcommand{\freq}{\nomUnc{60}{(1)}{}{Hz}}
\newcommand{\standAC}{\nomUnc{20}{(1)}{}{V}}
\newcommand{\varMin}{$3(3)\ \mathrm{V}$}
\newcommand{\varMax}{$123(3)\ \mathrm{V}$}

\newcommand{\transMin}{$12(2)\ \mathrm{V}$}
\newcommand{\transMax}{$1.14(6)\ \mathrm{kV}$}
\newcommand{\radiusMin}{$26.0\ \mu \mathrm{m}$}
\newcommand{\radiusMax}{$32.5\ \mu \mathrm{m}$}

\newcommand{\EscV}{$160(30)\ \mathrm{V}$}

\newcommand{\NullV}{$120(30)\ \mathrm{V}$}

\newcommand{\splitExpxr}{$21.5(1)\ \mathrm{mm}$}
\newcommand{\splitExpxl}{$-22.9(1)\ \mathrm{mm}$}

\newcommand{\splitComsolxr}{$21.1(0)\ \mathrm{mm}$}
\newcommand{\splitComsolxl}{$-22.0(0)\ \mathrm{mm}$}

\newcommand{\shuttleExpx}{$20.9(1)\ \mathrm{mm}$}
\newcommand{\shuttleComsolx}{$19.6(0)\ \mathrm{mm}$}

\DeclareMathOperator*{\argmin}{argmin}

\DTMnewdatestyle{slashdate}{%
}

\DTMsetdatestyle{slashdate}
\begin{document}
\title{An Accessible Planar Charged Particle Trap for Experiential Learning in Quantum Technologies}
\author{Robert E. Thomas}
\affiliation{Department of Physics, University of Washington, Seattle, Washington 98195, United States of America}
\affiliation{Department of Electrical \& Computer Engineering, University of Washington, Seattle, Washington 98195, United States of America}
\author{Cole E. Wolfram}
\affiliation{Department of Physics, University of Washington, Seattle, Washington 98195, United States of America}
\author{Noah B. Warren}
\affiliation{Department of Physics, University of Washington, Seattle, Washington 98195, United States of America}
\author{Isaac J. Fouch}
\affiliation{Department of Physics, University of Washington, Seattle, Washington 98195, United States of America}
\author{Boris B. Blinov}
\affiliation{Department of Physics, University of Washington, Seattle, Washington 98195, United States of America}
\author{Maxwell F. Parsons}
\email[]{mfpars@uw.edu}
\affiliation{Department of Electrical \& Computer Engineering, University of Washington, Seattle, Washington 98195, United States of America}

%Editor's Note: This paper presents the design and use of a charged particle surface trap, where all the electrodes are in a plane, an architecture commonly used in ion trap research. The authors demonstrate that macroscope-sized lycopodium spores, charged by the tribo-electric effect, can be trapped, shuttled, and split by their setup, providing a model system that allows students hands-on access to the technology underlying the promising use of ion traps for quantum computing. The trap design, key physics principles, and experimental procedures and results are presented, along with a computational method of data analysis. This work is appropriate for the undergraduate instructional laboratory and is accessible to those both inside and outside the fields of quantum information and AMO physics.

\begin{abstract}

We describe an inexpensive and accessible instructional setup that explores particle trapping with a planar linear ion trap. The planar trap is constructed using standard printed circuit board manufacturing and is designed to trap macroscopic charged particles in air. Trapping, shuttling, and splitting are demonstrated to students using these particles, which are visible to the naked eye. Students control trap voltages and can compare properties of particle motion with an analytic model of the trap using a computer vision program for particle tracking. Learning outcomes include understanding the design considerations for planar AC traps, mechanisms underpinning particle ejection,
the physics of micromotion, and methods of data analysis using standard computer vision libraries.
\end{abstract}

\maketitle

\section{Introduction} 
In recent years, rapid progress in controlling quantum states and entanglement between particles has paved the way for advancements in quantum computation, communication, and sensing.\cite{preskill, euro-roadmap-18} This ``second quantum revolution" is transitioning quantum science from academic research into industry-scale efforts, necessitating a corresponding expansion in quantum workforce development.\cite{workforce} To develop future quantum technologies, it is vital to bridge the gap between engineering and physics by exposing engineers to quantum-specific expertise and training physicists in engineering practices. Essential to this training is hands-on access to quantum hardware, enabling students to solidify classroom concepts in practical settings.

Trapped atomic ions are among the most effective qubit platforms for controlling and implementing quantum entanglement, demonstrating record entanglement fidelities and all-to-all qubit connectivity.\cite{understand_quant, progress_challenge, a_taxonomy} Quantum operations on trapped ion architectures typically involve the interaction of ions confined to one-dimensional Coulomb crystal structures.\cite{progress_challenge} Challenges arise as the atomic number density increases in these linear structures, including motional-mode crowding which hampers two-qubit gate fidelities and speeds.\cite{shuttle_based, beginner_guide, race-track, progress_challenge} One approach to balancing scalability and gate fidelity is the use of quantum charge-coupled devices (QCCDs).\cite{architecture} QCCDs aim for simultaneously high gate fidelities and scalability by separating long ion chains into smaller chains of higher gate fidelity through the individual and cumulative control of ion positions within the trap. QCCDs of the quadrupole trap variety, like ours, trap using a combination of static and oscillating electric fields.\cite{architecture, progress_challenge} With QCCDs, ion structures can be trapped in short chains and those chains can be split, merged, and shuttled with the manipulation of voltages on segmented DC electrodes.\cite{architecture, progress_challenge, shuttle_based} This approach enables the movement of pre-loaded ions from loading regions to interaction regions or to designated memory locations.\cite{architecture, shuttle_based} Despite their growing relevance for trapped ion quantum technology, workforce training on QCCD systems that trap atomic ions is challenging due to the need for integration of ultra-high vacuum systems, microfabricated traps, high-power radiofrequency signals, and complex laser systems for ion cooling, qubit manipulation, and qubit readout. By focusing solely on the physics and engineering of the planar trap, we can reduce both the complexity and cost of a lab where students can explore this technology. 

In this paper, we present an inexpensive, easily fabricated five-rail planar charged particle trap for use in physics and engineering laboratory courses to explore the basic functionalities of QCCD architectures. Our trap is designed on a standard printed circuit board (PCB) to trap negatively charged lycopodium moss spores, which we will refer to as ``particles,'' in air using static and $60\ \mathrm{Hz}$ oscillating electric fields. With this apparatus, students explore the fundamental principles of trapping charged particles, understand design considerations for these traps, examine the origin of specific trapped-ion dynamics like micromotion, and implement ion shuttling and splitting for positional control between trapping regions.

\section{Trapping Theory} 

\begin{figure}[htbp]
    \centering
    \includegraphics[width=\textwidth]{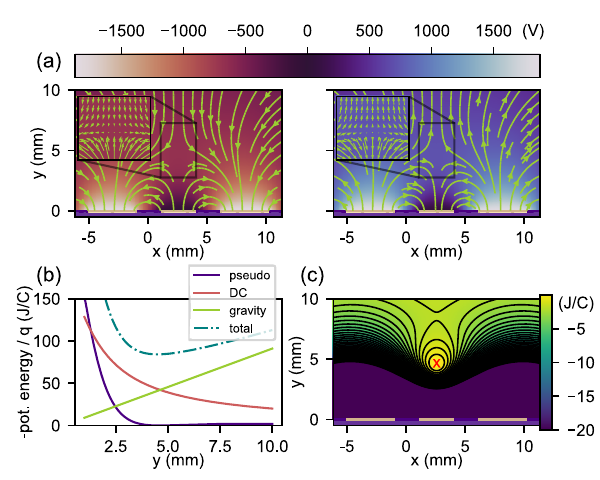} 
    \caption{(Color online) Model of the planar trap potentials.  Panel (a) shows the electric potential with field lines plotted at two instances of the AC potential that are $\pi$ radians out of phase.  The AC electrode voltages are at their minimum (left) and maximum (right). Panel (b) shows a \emph{y}-cut of the potential energy normalized by the charge with a breakdown of contributions to the trapping potential when the central DC electrode voltage is at 209 V. This voltage corresponds to alignment of the particle height with the AC null for a particle with charge-to-mass ratio of $-1.08\times10^{-3}$ C/kg.  Panel (c) shows only the pseudopotential energy from the AC electrodes, normalized by the charge, with the AC null labeled with \textcolor{red}{$\times$}.}  
    \label{fig: pseudo}
\end{figure}

Earnshaw's Theorem prohibits electrostatic systems from stably confining charged particles. To circumvent this obstacle, ion traps use combinations of DC ($\phi_{\mathrm{DC}}$) and AC ($\phi_{\mathrm{AC}}$) potentials for confinement.\cite{house, stable_analysis} At a given moment in time $t$, the potential in free space ($\phi$) at position $\mathbf{r} \equiv (x,\ y,\ z)$ can be written as
\begin{equation} \label{eq: dc+AC}
\phi(\mathbf{r}, t) = \phi_{\mathrm{DC}}(\mathbf{r}) + \phi_{\mathrm{AC}}(\mathbf{r})\cos{\Omega t}
\end{equation}
for angular frequency $\Omega$.\cite{house, stable_analysis} With the correct choice of trap parameters, one can indefinitely confine a charged particle within a small region of space through the time-averaged forces acting on the charged particle over an AC cycle. This method of confinement induces two characteristic forms of trapped-particle motion: micromotion and secular motion. Micromotion is oscillatory particle motion driven by the AC field, which oscillates at the AC field's oscillation frequency $\Omega$ at a relatively small amplitude.\cite{on_the} Secular motion corresponds to slower particle motion superimposed on the particle's micromotion. Properties of secular motion originate from the natural resonances of the effective trapping potential.\cite{on_the} This effective trapping potential is called the \emph{ponderomotive potential} or \emph{pseudopotential}, which can be interpreted as the trapped particle's potential energy function relative only to its secular
 motion dynamics. Thus, one can simplify the trapped particle's dynamics by approximating the time-dependent AC potential with the time independent pseudopotential using the following relationship:
\begin{equation}
\psi(\mathbf{r}) = \frac{q^{2}}{4m\Omega^{2}}|\nabla \phi_{\mathrm{AC}}(\mathbf{r})|^{2}
\end{equation}
where $q$ is the charge and $m$ is the mass of the particle.\cite{house}

The electric field from the AC electrodes at times $t = 0$ and $t = \frac{\pi}{\Omega}$ is shown in Fig.~\ref{fig: pseudo}(a). From the insets, one can see that the instantaneous field configurations can, at best, produce saddle-points of unstable equilibria instead of the local extrema necessary for stable equilibria. The time-averaged pseudopotential approximation gives a characteristic potential with a distinct local minimum as shown in Fig.~\ref{fig: pseudo}(c). With the addition of the linear gravitational potential [$\phi_{\mathrm{G}}(\mathbf{r}) \approx -gy$] and the potential from the central DC electrode [$\phi_{\mathrm{DC}}(\mathbf{r})$], one can approximate the cumulative, time-independent potential energy function $U(\mathbf{r})$ of an ion in the trap, normalized by the charge, as 
\begin{equation}
\frac{U(\mathbf{r})}{q}=  \frac{1}{\gamma}\phi_{\mathrm{G}}(\mathbf{r})+\phi_{\mathrm{DC}}(\mathbf{r})+\dfrac{\gamma}{4\Omega^2}|\nabla{\phi_{\mathrm{AC}}(\mathbf{r})}|^{2}
\end{equation}
where $\gamma$ is the charge-to-mass ratio $q/m$. The specific functional forms of $\phi_{\mathrm{AC}}$ and $\phi_{\mathrm{DC}}$ can be easily determined for the analytic model from geometric and voltage parameter values as described in the Appendix. The individual potential energy functions, in addition to  their sum along the unit normal of the central electrode $\hat{y}$ at $x = a/2$, is shown in Fig.~\ref{fig: pseudo}(b). Translating upward along this central electrode normal, the pseudopotential has a local minimum followed by a local maximum and then asymptotically approaches $0\ \mathrm{J / C}$. The local minimum of the pseudopotential in isolation is traditionally called the \emph{RF null}, which we call the \emph{AC null}, reflecting the 60 Hz operation of our trap. The local minimum of the total potential energy determines the particle's stationary height above the trap surface.

\section{Description of setup}

\begin{figure*}[htb]
    \centering
    \includegraphics[width=\textwidth]{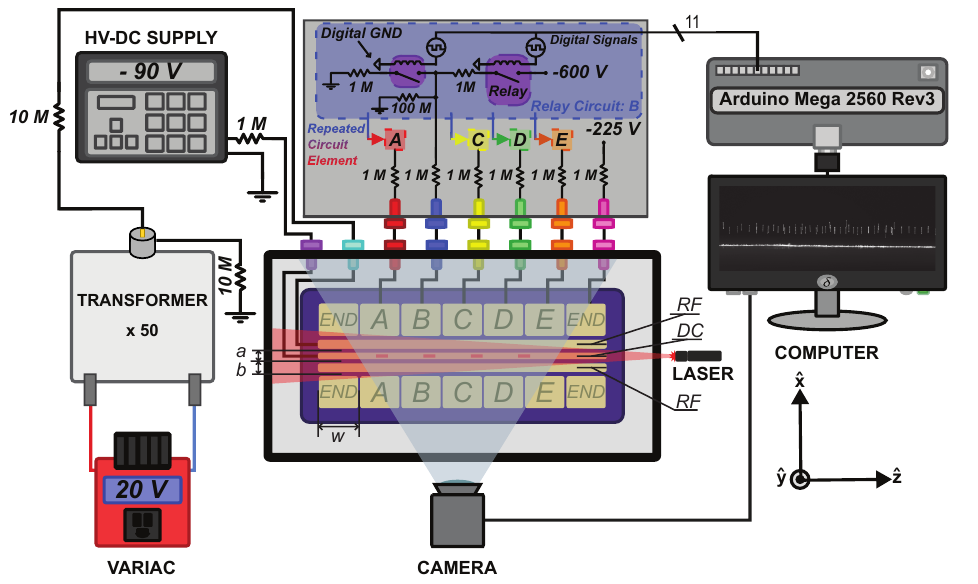}
    \hfill
    \caption{(Color online) The primary components of the lab are a Variac, step-up transformer, high voltage supply, enclosure, digital streaming camera, computer, relay circuit for  segmented electrode control, an Arduino, and a planar trap electrode configuration. The black-bordered rectangle represents our 3D-printed enclosure that surrounds the planar trap during operation to reduce turbulence and provide high-voltage protection. The main circuit elements for controlling the segmented shuttling electrodes are represented within the large gray rectangle above the enclosure. The input Variac provides a variable AC voltage input at \freq{} to a step-up transformer.  The transformer ground connects to the shared ground among the other high voltage supplies. The HV-DC supply operates from \centVmin{} to \centVmax{}. Two digital signal lines from the Arduino control the voltage on each of the 5 pairs of shuttling electrodes, labeled $A$, $B$, $C$, $D$, and $E$ for a total of 10 signal lines with 1 shared ground (11 lines). A 635 $\mathrm{nm}$ wavelength diode laser illuminates the trapped particles, with scattered light recorded by the camera that sends video data to a computer. The computer stores data, analyzes data through our Python code, and controls the Arduino for user controlled, coordinated switching of shuttling electrode voltages. The experimental trap dimensions that are relevant to the analytic model in Sec.~III are labeled $a$ and $b$, and $w$. The endcap electrodes, correspondingly labeled \textit{END} are controlled with a high voltage DC-DC converter providing a maximum voltage of $-2.441(32)\times 10^{2}\ \mathrm{V}$.}
    
    \label{fig: layout}
\end{figure*}
We developed a planar quadrupole trap for macroscopic particles at atmospheric pressure, which accepts user
inputs to control charged-particle confinement, shuttling and splitting. This trap, shown in Fig.~\ref{fig: layout}, is manufactured on a standard PCB and consists of five rails with segmented, individually addressable electrodes that control in-trap shuttling in a similar fashion to traps of Dr. Kenneth Brown \textit{et al}.\cite{pcb_trap} High voltage reed relays actuated via an Arduino microcontroller configure a set of voltage dividers to provide each shuttling electrode with one of three programmable voltages. The setup features a 3D printed enclosure for user safety and turbulence reduction, an illumination laser, and a camera for imaging. Video data is processed with the OpenCV Python package. Image processing code, CAD models, Gerber files, and data analysis/visualization scripts are publicly accessible.\cite{github} A full parts list can be downloaded as supplementary materials to this paper.

\noindent \textbf{Charged Particles:}
We trap charged lycopodium moss spores which have a diameter of \radiusMin{}\cite{yellow1, chaos} to \radiusMax{} \cite{timmer} and are visible to the naked eye with adequate illumination. A polytetrafluoroethylene (PTFE) wand and a wool cloth charge these particles through the triboelectric effect \cite{yellow1, chaos} to charge-to-mass ratios ranging from 
$2 \times 10^{-4}\ \mathrm{C/kg}$\cite{chaos} to
$5 \times 10^{-2}\ \mathrm{C/kg}$.\cite{timmer}
The PTFE rod serves the dual purpose of charging the particles, while shielding the user from the high voltages of the planar trap's electrodes.\cite{chaos} Similar macroscopic traps used polyethylene microspheres,\cite{chaos} glass spheres,\cite{chaos, stable_structure} stainless steel spheres,\cite{chaos} aluminum spheres,\cite{macro} aluminum oxide spheres,\cite{stable_structure} anthracene dust particles,\cite{simple} glycerin oil droplets, \cite{oil} and borate particles, among other materials.\cite{macro}  We use lycopodium moss spores because they are inexpensive, monodispersed, roughly spherical, non-toxic, easily chargeable, and non-conductive.\cite{yellow1}

\noindent \textbf{Planar Trap Geometry:}
Our electrode geometry is of the five-rail variety, in part due to its popularity in industry because of its compatibility with standard microfabrication techniques. \cite{architecture, guide, littich, progress_challenge}
The geometry consists of a central DC electrode of dimensions \dimCentw{}  by \dimCentl{}, two high voltage AC electrodes of dimensions \dimACw{} by \dimACl{}, and two  rows of seven segmented DC electrodes of dimensions \dimSegw{} by \dimSegl{}. The dielectric separation between the central DC and high voltage AC electrode is \ACGNDspace{}, the separation between the AC and segmented electrodes is \segACspace{}, and the separation between segmented electrodes is \segSegspace{}. Every electrode has $1.0(1)$ mm corner radius fillets to prevent arcing.\cite{proto_express} The two AC electrodes are shorted, and each segmented electrode on the top row of the trap is directly linked to the segmented electrodes on the bottom row along the vertical. These shuttle-electrode pairs are labeled \textit{A}, \textit{B}, \textit{C}, \textit{D}, and \textit{E}. The four endcap electrodes at each corner are connected together. The planar trap was manufactured on a two-layer PCB with FR408HR 
as its insulating core with a dielectric strength of $70\ \mathrm{kV/mm}$.

\noindent\textbf{Circuits:}
A Variac supplies low AC voltages which are stepped up by a transformer from an input voltage range of \varMin{} to \varMax{} to an output voltage range of \transMin{} to \transMax{}. The HV-AC output is connected to the AC electrodes through a $10\ \mathrm{M}\Omega$ resistor, while the ground reference of the transformer is shared among the other high voltage supplies, including a high voltage DC-DC converter that sources HV-DC for the endcap electrodes. The other leg is connected to the two high-voltage AC electrodes. A $10\ \mathrm{M}\Omega$ resistor between the HV output and ground limits the current in the event of a short.\cite{yellow1, chaos}
The segmented electrode driver circuit is supplied $-600(1)\ \mathrm{V}$, which is fed into a network of voltage dividers, providing controllable output voltages of \highV{} (\texttt{high}), \lowV{} (\texttt{low}), and \offV{} (\texttt{off}) with relay switching times of \relayT{}. Each output has a corresponding $1\ \mathrm{M}\Omega$ resistor before connecting to each segmented electrode pair to limit currents from incidental shorts from the suface electrodes. The central DC electrode has an applied voltage within the experimentally relevant range of \centVmin{} to \centVmax{}, which is sourced from a Stanford Research Systems Model PS350 high voltage DC power supply. We calibrate voltages on the electrodes by measuring them directly at the electrode surface using a high-voltage probe. 

The planar trap is housed in an enclosure composed of polylactic acid (PLA) with acrylic optical ports on two perpendicular faces of the enclosure for the optical illumination and imaging of the trapped particles. The enclosure secures the planar trap in place, reduces air turbulence, houses the high voltage wires, and provides a physical barrier between the user and the high voltage electrodes. It was designed to be printed on a 3D printer with minimum bay space of dimensions $255\ \mathrm{mm}$ by $155\ \mathrm{mm}$ by $170\ \mathrm{mm}$.

\noindent\textbf{Imaging and Image Processing:}
A $1\ \mathrm{mW}$, $635\ \mathrm{nm}$ wavelength laser is used to illuminate the particles, which are imaged onto a FLIR Blackfly S monochrome camera with a maximum frame rate of $60$ frames per second and a resolution of 1,616 by 1,240 pixels. A $6\ \mathrm{mm}$ focal length lens is used with an aperture of f/1.85. For automated data analysis, we use image detection methods from the OpenCV Python package to log the positions of the trapped particles within recorded video files.\cite{blob}  This method incorporates a thresholding algorithm, followed by filtering of resulting connected regions by size to isolate trapped ions from background noise.

\section{Laboratory exercises}
\subsection{Measuring charge-to-mass ratio and exploring motion of a single trapped particle}\label{sec:micromotion}\label{sec:gamma}

The charge-to-mass ratio $\gamma$ of a  charged particle confined within the trap is an important quantity for determining and optimizing the particle's trapping stability. Provided below are two experiments using the planar trap in which students experimentally determine $\gamma$ for a particle within the trap. Students can compare experimental properties of the five-rail trap with an analytic model developed by M.G. House.\cite{house}  See the Appendix for an explicit formulation of the trap potentials for our apparatus.\\

\noindent\textbf{Method 1, Fitting $\gamma$ to height vs. central DC electrode voltages:}
For stable, continuous particle confinement above the trap, the particle must be positioned at a local minimum of $U$. Let this local minimum be expressed as $\mathbf{r}_{\mathrm{min}} \equiv (a/2,\ y_{\mathrm{min}})$ for some $y_{\mathrm{min}} > 0$ with the $z$-coordinate being neglected due to the trap's translational symmetry along the $z$-axis. Therefore
$-\nabla U(\mathbf{\mathbf{r}_{\mathrm{min}}}) = \mathbf{0}$.
Inspection of the total potential in Fig. \ref{fig: pseudo}(b) suggests that there exists a unique minimum $y_{\mathrm{min}}$ for $y>0$. Thus

\begin{equation}
\begin{split}
y_{\mathrm{min}} =\argmin_{y>0} \Big\{\frac{1}{\gamma}\phi_{\mathrm{G}}(y)+\phi_{\mathrm{DC}}(y)+\\\dfrac{\gamma}{4\Omega^2}|\nabla{\phi_{\mathrm{AC}}(y)}|^{2}\Big\}
\end{split}
\end{equation}

\noindent with the knowledge that $x = a/2$ is now temporarily a set parameter of $\phi_{i}$. One can determine $\gamma$ experimentally by measuring particle height values with different applied DC voltages and fitting those data to corresponding $y_{\mathrm{min}}$ points of the analytic model at the same applied central DC voltages. \\

\noindent\textbf{Method 2, Determining $\gamma$ from minimal micromotion amplitude:}
The micromotion amplitude $\alpha$, to first order, is related to the displacement from the AC null by $\alpha \propto |y_{\mathrm{null}} - y|$.\cite{pearson_mast} Therefore, one can experimentally determine the AC null position from the stationary height of the particle above the trap corresponding to the smallest measured micromotion amplitude. We will consider a charged particle that is stably trapped at the AC null, $\mathbf{r}_{\mathrm{null}} \equiv (a/2,\ y_{\mathrm{null}})$. A stably trapped particle implies that \textcolor{black}{$-\nabla{U(\mathbf{r}_{\mathrm{null}})} = \mathbf{0}$}. Therefore
\begin{equation} -mg -q\nabla{\phi_{\mathrm{DC}}(\mathbf{r}_{\mathrm{null}})}-\frac{q^{2}}{4m\Omega^2}\nabla|\nabla{\phi_{\mathrm{AC}}(\mathbf{r}_{\mathrm{null}})}|^{2} = \mathbf{0}
\end{equation}
 Since the AC null location is defined as a local minimum of the AC pseudopotential, \textcolor{black}{$(q^{2}/4m\Omega^2)\nabla\big(|\nabla{\phi_{\mathrm{AC}}(\mathbf{r}_{\mathrm{null}})}|^{2}\big) = \mathbf{0}$}. When a particle sits at the AC null, the vertical force from the \textcolor{black}{central} DC electrodes balances gravity, i.e. 
$$-mg-q\nabla \phi_{\mathrm{DC}}(\mathbf{r}_{\mathrm{null}}) = \mathbf{0}$$
\begin{equation} \label{gamma}
\Longrightarrow \gamma = \frac{q}{m} =\dfrac{-g}{\big|\nabla{\phi_{\mathrm{DC}}(\mathbf{r}_{\mathrm{null}}})\big|}.\end{equation}

\textcolor{black}{The location $y_\mathrm{min}$, as described in Method 1 above, is the total potential energy function's local minimum}, the centroid of oscillation of the ion's micromotion in the trap.  By minimizing micromotion, one can gauge the closeness of the particle's height $y$ relative to the AC null location $y_{\mathrm{null}}$ with the relationship $\alpha \propto |y_{\mathrm{null}} - y|$. Once the local minimum position is found, one can approximate $y_{\mathrm{min}} \approx y_{\mathrm{null}}$ which, with Eq.~(\ref{gamma}), allows one to calculate $\nabla \phi_{\mathrm{DC}}$ at $\mathbf{r}_{\mathrm{null}} = (a/2,\ y_{\mathrm{null}})$ and thus determine $\gamma$.  

\begin{figure*}[htbp]
    \centering
    \includegraphics[width=0.66\textwidth]{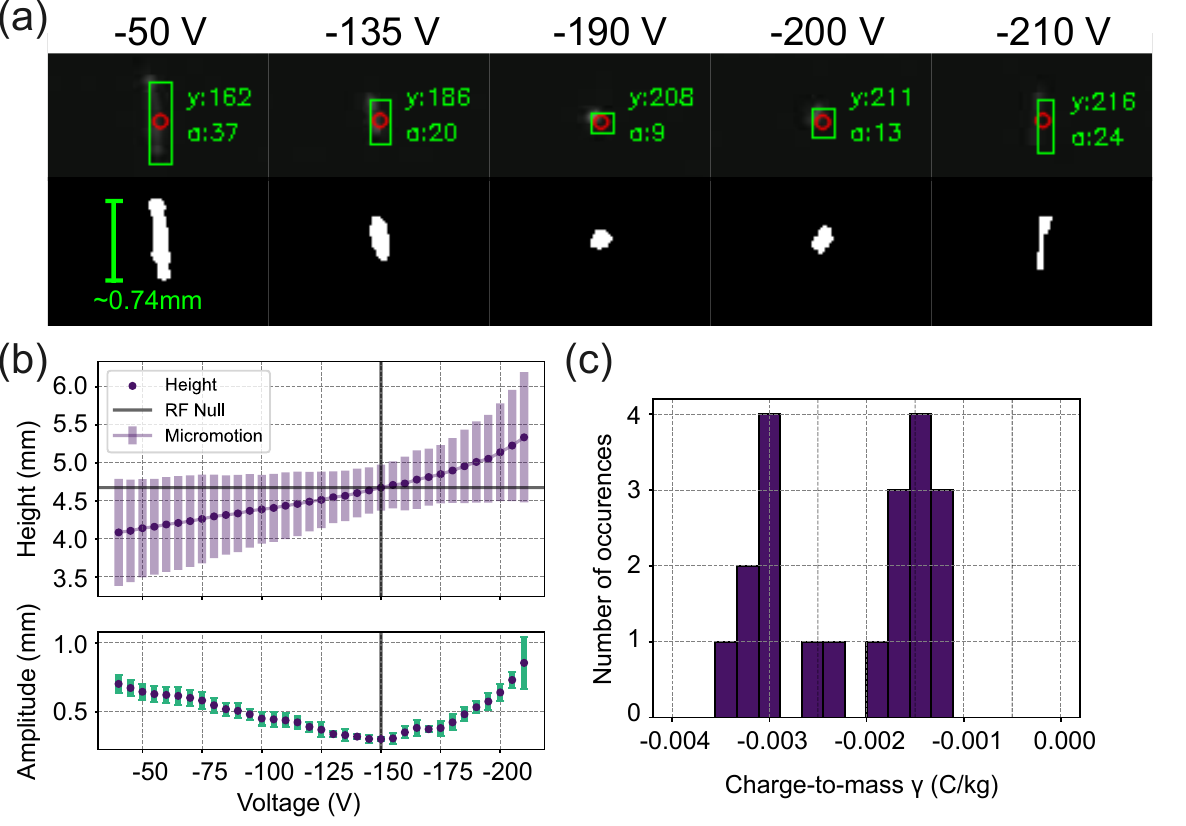}
    \caption{(Color online) Panel (a) shows representative images of a detected particle at different central electrode voltages with the location and amplitude of the particle labeled in units of pixels. On the bottom of Panel (a) is the binarized image used to better determine particle position and micromotion amplitude. The top axis of Panel (b) shows the height of the particle versus the central DC electrode voltage with the micromotion amplitude indicated by the light purple bars. The micromotion amplitude versus voltage is shown on the bottom axis, with error bars showing the standard deviation of the micromotion amplitude over 15 frames at the same voltage. The AC null point is indicated on both graphs with the gray crosshair. Panel (c) is a histogram of the calculated charge-to-mass ratios $\gamma$ of an ensemble of 20 different trapped particles.}
    \label{fig:micromotion}
\end{figure*}

\noindent\textbf{Experimental Methods:}
We image the particles with a 6 mm focal length lens and calibrate the distance per pixel units in the video by imaging a fine ruler vertically above the central DC electrode. For loading charged particles into the trap, the shuttling electrodes \textit{A}-\textit{E} are grounded, the Variac is set to \standAC{}, and the endcap electrodes are set to -244(3) V. We charge and deposit an ensemble of particles into the trapping region by tapping the \textcolor{black}{PTFE} rod dipped in lycopodium spores above the trap through a hole in the enclosure. To isolate a single particle, we lower the AC trapping voltage slowly until only a single particle remains in the trap, effectively filtering out the ions with lower charge-to-mass ratios. Then, the Variac is returned to \standAC{}. The motion of the particle is captured in a video, with the experiment starting with $-40(1)\  \mathrm{V}$ on the central electrode. This is incrementally decreased by $-5(1)\  \mathrm{V}$ every 5 seconds until the particle is ejected from the trap, in which case the video recording is terminated. 

The data analysis code uses OpenCV to identify the location and dimension of the particle's motion throughout the video, as shown in Fig.~\ref{fig:micromotion}(a). Increasing the exposure time of the camera causes the illumination of the particle's complete trajectory, which appears as a ``streak'' in the video and allows for measurement of micromotion amplitude. We operate the camera at a fixed frame rate and determine time from the frame number. After collecting the height and micromotion values over a number of user-defined frames (15, in our case), the averages and standard deviations of the centroids and the micromotion amplitudes of the particles are output to that trial's data set. The code repeats this process for each data point, continuing until the particle is no longer visible (i.e., ejected from the trap).

\begin{figure}[htbp]
\includegraphics[width=0.66\textwidth]{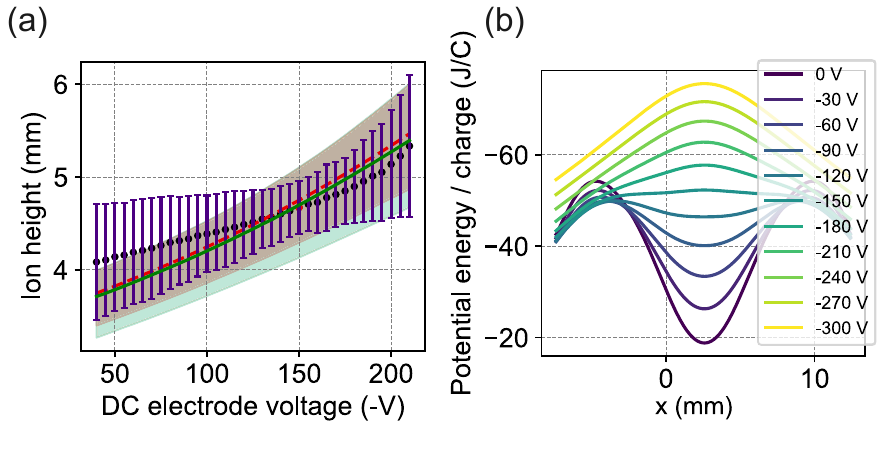}
        \caption{(Color online) Panel (a) shows particle height as a function of applied central DC voltage for a specific trapped particle. Methods 1 and 2 for determining the charge-to-mass ratio are applied to this data set.  The dashed red line and red band correspond to the best-fit and error bars from Method 1.  The solid green line and green band represent the best-fit and one-standard-deviation error from Method 2, with the red data point near 150 V indicating the point with minimal micromotion used to calculate the charge to mass ratio.  The charge-to-mass ratios calculated from Methods 1 and 2 are \chargeMassOne{} and \chargeMassTwo{}, respectively.  The errors are dominated by errors in determining the position of the particle. Panel (b) shows the potential along the $x$-axis at various applied central DC voltages with an example particle of charge-to-mass ratio of $-1.6\times10^{-3}$ C/kg. 
 The potentials are symmetric about the trap center.}
        \label{fig: fitting}
\end{figure}

\noindent\textbf{Results and Discussion:}
Over 20 independent trials, particle height data, in addition to micromotion amplitude length, were saved as a function of central electrode voltage starting at $-40\ \mathrm{V}$ and decreasing by $-5\ \mathrm{V}$ intervals until particle ejection. These data provide the opportunity to apply both Methods 1 and 2 towards calculating the same particle's charge-to-mass ratio, which can be used as the parameter values in $U / q$ to create fitted or extrapolated position curves as a function of central electrode voltage. A representative example is provided in Fig. \ref{fig: fitting}.

Applying Methods 1 and 2 to our positional data, one obtains \chargeMassOne{} and \chargeMassTwo{} with respective standard deviations of \stdCmone{} and \stdCmTwo{}. Inputting these values back into $U/q$ as parameters produces the green extrapolated position curve for $\gamma_1$ and the red fitted positions curve for $\gamma_2$ with corresponding reduced chi-squared values of \rChione{}\ and \rChitwo{}. Since $1 < \hat{\chi}^{2}_{1} < \hat{\chi}^{2}_{2} < 5$, the goodness-of-fit test suggests that the analytic planar trap model, in conjunction with Methods 1 and 2, produces an accurate predictive model. Uncertainty in Method 1's charge-to-mass ratio ($\delta \gamma_{1}$) in addition to reduced chi-squared values $\chi_{1}^{2}$ and $\chi_2^{2}$ were calculated with the method of generalized linear least-squares fitting.\cite{fitt_book} 

We determined $\gamma$ for all 20 trials using Method 2, as shown in Fig. \ref{fig:micromotion}. The mean charge-to-mass ratio of the particle ensemble is \chargeMassmean{} with a standard deviation of \stdcm{} C/kg. These data are in agreement with the charge-to-mass ratios of  $-2 \times 10^{-4}\ \mathrm{C/kg}$\cite{yellow1} to $-5 \times 10^{-2}\  \mathrm{C/kg}$\cite{chaos, timmer} cited in the literature. From the micromotion data over all trials, the mean AC null height was calculated to be \meanACnull{} above the surface of the trap. The experimentally derived AC null height is in agreement with the analytic model's AC null height of \modelACnull{}, with corresponding uncertainty being solely from the uncertainties in geometric values $a$, $b$, and $g$ of the manufactured board.\cite{house} 

The average voltage required to minimize each particle's micromotion is \NullV{} and the average voltage to cause particle ejection from the trap is \EscV{}. A visual of the potential energy along the \emph{x}-axis at the AC null height for a particle of $\gamma = -1.63(40)\times10^{-3}$ C/kg is shown in Fig. \ref{fig: fitting}(b). Our mean values match the physical intuition implied from the potential curvature, as the voltage range $-120\ \mathrm{V}$ to $-180\ \mathrm{V}$ corresponds to a transition from confinement to anti-confinement along the \emph{x}-axis. 

\subsection{Ion Shuttling and Splitting}
The primary appeal of QCCDs is dynamical positional control over trapped ions via segmented DC electrodes. Our apparatus allows students to explore the dynamics of  shuttling and splitting ensembles of particles in these devices.   

\noindent\textbf{Experimental Methods:}
Distance calibrations are conducted by placing a fine ruler horizontally, flush along the central electrode. A calibration image is then taken, which is used to determine the pixel-to-millimeter conversion factor. For both the shuttling and splitting experiments, the \textcolor{black}{Variac} is set to $20(3)\ \mathrm{V}$ RMS which is stepped up to an RMS voltage of $963(3)\ \mathrm{V}$. The shuttling DC electrodes are supplied $-600\ \mathrm{V}$, which is voltage-divided into the \texttt{off}, \texttt{low}, and \texttt{high} voltage settings through the relay network shown in Fig. \ref{fig: layout}. The end caps are supplied with $-2.441(32)\times 10^{2}\ \mathrm{V}$ and the central DC electrode is grounded.

To start each experiment, we load the trap with an ensemble of particles as in Sec.~\ref{sec:gamma}. The particles are positioned at the center of the trapping axis within the neighborhood of the \textit{C} DC electrode pair by assigning the voltage value of high on electrodes \textit{A} and \textit{E}, low on electrodes \textit{B} and \textit{D}, and off on electrode \textit{C}. Once centered, we filter out those with lower charge-to-mass ratios until only one or two particles remain in the trapping region to conduct shuttling and splitting, respectively. 

\begin{figure}[htbp]
        \includegraphics[width=0.66\textwidth]{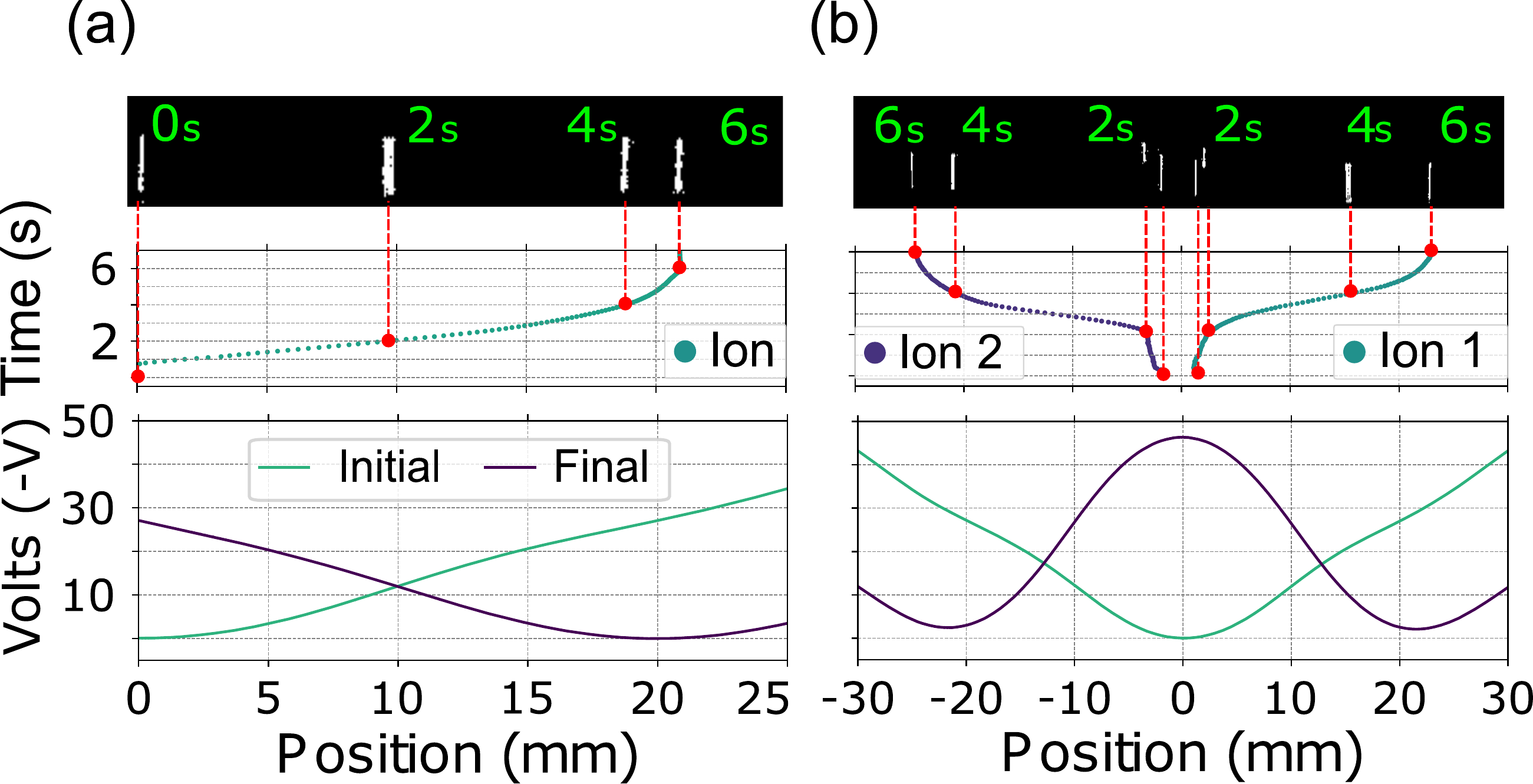}
        \caption{(Color online) Panel (a) shows charged-particle shuttling and Panel (b) shows particle splitting. In both panels, the top images show snapshots of charged particle positions in the trap as seen by our tracking program. These images correspond to the red highlighted data points in the middle plots, where the full set of ion position data is extracted with our tracking program. The initial and final particle positions are controlled by the locations of the initial and final potential pattern minima. The bottom graphs show each experiment's corresponding COMSOL finite-element simulations of the potential from the segmented electrodes at the height of the analytic model's predicted AC null. The green potential lines in these potential plots represent the initial potential while the dark purple lines represent the final potential. For shuttling, the distance between the potential minima of the initial and final potential pattern minima is \shuttleComsolx{} in the 
COMSOL simulation. The experimental shuttling distance was \shuttleExpx{}. For splitting, the distance traveled between the two particles' initial and final positions is simulated to be \splitComsolxl{} and \splitComsolxr{} while the experimental distances were measured to be \splitExpxl{} and \splitExpxr{} respectively. Note that the position axis here corresponds to the z-axis of Fig.~\ref{fig: layout}.}
        \label{fig: shuttling}
\end{figure}

In both experiments, we switch the voltage configuration of the segmented electrodes and observe the resulting transient motion of the particle. For shuttling, the initial potential pattern with the trap centered on \textit{C} is switched to a final pattern where the trap is centered on \textit{D}: high on electrodes \textit{A} and \textit{B}, low on electrodes \textit{C} and \textit{E}, off on electrode \textit{D}, and $-$244(3) V on the \textit{END} electrodes. For splitting, the same initial potential pattern is switched to a final pattern that creates a double well potential around electrode \textit{C} in two stages. First, all of the segmented DC electrodes are set to \texttt{off}. Then, electrode \textit{C} is set to \texttt{high}, electrodes \textit{A} and \textit{E} are set to \texttt{low}, and electrodes \textit{B} and \textit{D} are set to \texttt{off} to create a double-well potential as shown in Fig. \ref{fig: shuttling}(b). Each shuttling and splitting video is then analyzed with OpenCV in the same fashion as the micromotion experiment. 

\noindent\textbf{Results and Discussion:}
Fig.~\ref{fig: shuttling}(a) and \ref{fig: shuttling}(b) show data obtained for shuttling and splitting, respectively. Comparing these data to corresponding COMSOL simulations of the potential curvature along the AC null during each process, one can see that the experimental results map well to the COMSOL simulation. 
The end locations for the particles in each experiment also correspond to the positions of the local minima of the potentials.   

\section{Extensions to the setup}
We envision a number of additional exercises made possible with our setup, which can extend the laboratory experiments described in this paper. One follow-up design exercise is to imagine how the setup must be modified to trap Ca$^{+}$ ions, where the charge-to-mass ratio is $2.4 \times 10^{6}\ \mathrm{C/kg}$. Electrical engineers can be asked to identify design solutions that produce the needed voltages and frequencies for optimal trapping and shuttling performance. Another simple hands-on exercise is to measure the micromotion frequency of the trapped particles by either strobing the illumination laser at different frequencies or sweeping the camera frame rates, both using the stroboscopic effect.\cite{yellow1} Students can also determine the damping coefficient of air based on the damped dynamics of the shuttling particles\cite{pearson_mast} by fitting the particle trajectory data to a mechanical model with damping, either extending the analytic model of the trap to includ
e the segmented electrodes or using the COMSOL model from Fig.~\ref{fig: shuttling}.  Exploring the COMSOL finite-element model further, perhaps through comparison to the analytical model we present, is an opportunity to introduce students to finite-element methods.  The transition between 1D and 2D Coulomb crystals is another property students can explore with adequately large segmented electrode voltages.\cite{yellow1}

There are several modifications to the apparatus that will improve or expand the capabilities of the lab if implemented. The addition of a grounded plate above the trap will increase trap depth and improve control over fields from stray charges.{\cite{pcb_trap, pearson_mast}} Distinct particle-chain secular mode excitation may be possible on this setup with the usage of electrospray ionization with glycerin. \cite{oil}  We have been unable to observe clearly resolved secular modes with the lycopodium moss spores, most likely because these modes are in the overdamped regime.\cite{chaos, simple} Another possible improvement is to apply analog signals to the shuttling electrodes via a digital-to-analog converter, allowing for smoother shuttling, the exploration of different shuttling waveforms and their impact on shuttling and splitting.\cite{on_the} The setup can be taken to the next level of relevance to current research by redesigning it to implement``X'' and ``T'' junctions,
 which are important in QCCDs.\cite{architect}  More exotic AC trap geometries, such as 2D arrays\cite{pearson_art} or 3D multilevel traps\cite{on_the} would also be interesting to explore in a macroscopic regime. 

\section{Acknowledgements}
We thank Enrique Garcia and Jaidon Lybbert for their invaluable guidance with the relay circuit design. We thank Dr. Alejandro Garcia for loaning us a high-voltage power supply early in the project. We thank Carl Thomas, Jane Gunnell, and Dr. Alexander Kato for their assistance, and Dr. Sara Mouradian for her suggestions on determining the necessary shuttling voltages and feedback on the manuscript.  We thank Chukwuemueka Mordi for his feedback on the accompanying code.
This work was supported by the National Science Foundation under awards DMR-2019444 (IMOD an NSF-STC) and PHY-2011503, and by the University of Washington Student Technology Fee.  This project is part of the Quantum Technologies Training and Testbed (QT3) laboratory at the University of Washington, which is supported by NIST award 60NANB23D202.

\noindent\textbf{Data Availability Statement:}
The design data and analysis code that supports the findings of this study are openly available in \texttt{qt3-ion-trap} at \url{https://github.com/qt3uw/qt3-ion-trap.git}.\cite{github}  The raw video data is available upon request.

\section*{Appendix: Analytic Model Free-Space Potential Functions} \label{appen: b}
We can use an analytic model to produce specific functions for $\phi_{\mathrm{AC}}(\mathbf{r})$ and $\phi_{\mathrm{DC}}(\mathbf{r})$. Let the origin of our coordinate system be at the bottom left-most corner of the central electrode.  We used the following approximations in our analytic model of the trap: the widths of the central and AC electrodes extend infinitely along the linear trapping axis; the lengths of the segmented electrodes extend infinitely away from the trapping region; each electrode is rectangular; the electrodes are perfect conductors with uniform voltage; the potential goes to zero as you approach infinity in the direction normal to the trap surface.\cite{house}  A finite volume of dielectric material is necessary between electrodes to prevent arcing and dielectric breakdown between neighboring electrodes. To account for this perturbation on our analytic model, we implemented House's recommendation of a linear voltage interpolation between the voltage value
s of neighboring electrodes applied in the inter-electrode dielectric regions.\cite{house} We compared this interpolation to  approximating the dielectric gaps to be infinitesimal and found better experimental agreement in particle height versus central DC electrode voltage using the linear interpolation. Then, an electrode with edges at locations $x_{1}$ and $x_{2}$ relative to the origin has an in-plane voltage distribution of 
$$\phi_{i}(x, y=0) = 
    \begin{cases}
        V_{i},\ x \in (x_{1}, x_{2}) \\
        0, \text{ else}
    \end{cases}$$
which produces a corresponding free-space voltage above the plane of $\phi_{i}(x, y > 0) =$

\begin{equation}
\frac{V_{i}}{\pi} \Big[\arctan\big(\frac{x_{2} - x}{y}\big) - \arctan\big(\frac{x_{1} - x}{y}\big)\Big]
\end{equation} where the potential is independent of the z-coordinate due to the translational symmetry of the infinite electrode length assumption.\cite{house} Therefore, a central electrode of sides $x_{1} = 0, x_{2} = a$ has an in-plane voltage of
$$\phi_{\mathrm{DC}}(x, y=0) = 
    \begin{cases}
        V_{\mathrm{DC}},\ x \in (0, a) \\
        0, \text{else}
    \end{cases}$$
with corresponding free-space voltage of $\phi_{\mathrm{DC}}(x, y > 0) = $
\begin{equation} \label{eq: dc}
\frac{V_{\mathrm{DC}}}{\pi}\Big[\arctan\big(\frac{a - x}{y}\big) + \arctan\big(\frac{x}{y}\big)\Big]
\end{equation}
and AC electrodes with
$$
    \phi_{\mathrm{AC}}(x, y = 0, t) = 
    \begin{cases}
        V_{\mathrm{AC}}\cos{\Omega t},\ x \in (-c, 0) \cup (a, a + b) \\
        0,\ \text{else}
    \end{cases}$$
producing free space voltages of $\Phi_{\mathrm{AC}}(x, y > 0, t) =$ $$ \frac{V_{\mathrm{AC}}}{\pi}\Biggr[\arctan\big(\frac{a + b - x}{y}\big) + \arctan\big(\frac{c + x}{y}\big)$$  $$- \Big(\arctan\big(\frac{a-x}{y}\big) + \arctan\big(\frac{x}{y}\big)\Big)\Biggr]\cos{\Omega t}$$
\begin{equation} \label{eq: AC}       
 \Longrightarrow  \Phi_{\mathrm{AC}}(x, y > 0, t) = \phi_{\mathrm{AC}}(x, y > 0)\cos(\Omega t)
\end{equation} where $\phi_{\mathrm{AC}}(x, y > 0)$ is the separable, time independent part of $\Phi_{\mathrm{AC}}(x, y > 0, t)$. Therefore, we can combine equations ~\ref{eq: dc} and ~\ref{eq: AC} get a solution for $\frac{U}{q}$.

One can determine $\frac{U}{q}$ without the infinite electrode assumption by using the analytic function for a finite electrode described by House in addition to simulating other finite electrodes in the model such as the endcap electrodes. \cite{house, fisher} However, calculating $|\nabla \phi_{\mathrm{AC}}(\mathbf{r})|^2$ becomes significantly more difficult. 

\begin{singlespace}
\clearpage
\onecolumngrid
\section*{Bill of Materials for Planar Particle Trap}
\begin{table*}[h]\centering
\begin{tabular}{|l|}
\hline 
\textbf{Total Apparatus Price: \$6,818.17 (USD as of October, 2024)} \\
\hline
\end{tabular}
\end{table*}
\begin{table*}[!htb]\centering
\caption{Primary Components and Breadboard Peripherals}\label{tab: primary}
\scriptsize
\begin{tabular}{||l r r r r r r|} \toprule
\hline
\textbf{Item} & \textbf{Application} & \textbf{Supplier} & \textbf{Part Number} & \textbf{Price/Item} & \textbf{Qty.} & \textbf{Price} \\ 
\hline\hline \multicolumn{7}{|c|}{\textbf{Primary Components}} \\ \hline
Variac & AC Voltage Supply &Amazon & B07Y4ZY3K1 &\$57.17 &1 &\$57.17 \\
Transformer &Step-up AC Voltage &Static Clean&F-90117 &\$437.00 &1 &\$437.00 \\
HV Cable &Transformer Peripheral &Static Clean &F-80312 &\$11.00 &1 &\$11.00 \\
HP Connector Kit &Transformer Peripheral  &Static Clean&F-90100 &\$20.00 &1 &\$20.00 \\
HV-DC Supply &DC Voltage Supply & SRS\footnote{Stanford Research Systems} &PS310 &\$2,650.00 &1 &\$2,650.00 \\
HV-DC Supply &Endcap Voltage Supply &Digikey &1470-3186-ND &\$138.81 &1 &\$138.81 \\
Function Generator &Produce END Voltage &Siglent &SDG1032X &\$359.00 &1 &\$359.00 \\
Arduino Mega & Control of Reed Relays &arduino.cc &A000067 &\$53.00 &1 &\$53.00 \\
Blackfly S Camera &Imaging &Edmund Optics &22-078 &\$655.00 &1 &\$655.00 \\
1 mW Laser &Illumination &Edmund Optics &37-029 &\$123.00 &1 &\$123.00 \\
PLA (1.75 mm, Black) &Structural Support &Amazon & B07PGY2JP1 &\$23.99 &3 &\$71.97 \\
Acrylic &Optical Ports &Amazon & B07RY4X9L3 &\$16.95 &1 &\$16.95 \\
Breadboard &Fastening Components &Thorlabs &MB1218U &\$216.05 &1 &\$216.05 \\
6 mm Lens &Camera Lens &Edmund Optics &33-301 &\$260.00 &1 &\$260.00 \\
% 25 mm Lens &Micromotion Lens &Edmund Optics &59-871 &\$325.00 &1 &\$325.00 \\
Wool &Particle Charging &Amazon &B07D6XG3S4 &\$9.89 &1 &\$9.89 \\
Lycopodium Moss Spores &Working Ions &SKS\footnote{\url{<sciencekitstore.com>}} &LYCOP2 &\$69.00 &1 &\$69.00 \\
\textcolor{black}{Polytetrafluoroethylene (PTFE)} Rod &Particle Charging &Amazon &B01BSLI1YG &\$12.03 &1 &\$12.03 \\
Cat 5e GigE Cable &Camera Signal &Edmund Optics&86-785& \$62.00& 1 & \$62.00 \\
\hline \multicolumn{7}{|c|}{\textbf{Breadboard Peripherals}} \\ \hline
Laser Diode Mount &Fasten Laser &Edmund Optics &37-144 &\$90.00 &1 &\$90.00 \\
Camera Mount &Fasten Camera &Edmund Optics &15-838 &\$37.50 &1 &\$37.50 \\
Switch &Toggle Laser &Digikey &360-1887-ND &\$6.70 &1 &\$6.70 \\
Laser Power Supply &Power Laser &Edmund Optics &59-099 &\$66.50 &1 &\$66.50 \\
Metal Screws and Bolts &Fastening Breadboard Items &Thorlabs &HW-KIT2 &\$139.58 &1 &\$139.58 \\
Post Clamps &Fastening Laser and Camera &Thorlabs &CF125 &\$9.68 &2 &\$19.36 \\
Post Holders &Fasten Posts to Breadboard &Thorlabs &PH1.5E &\$27.92 &2 &\$55.84 \\
1.5'' Posts &To hold Camera and Laser &Thorlabs &TR1.5 &\$5.65 &2 &\$11.30 \\
Nylon Screw (6-40) &Fasten Trap to Enclosure &McMaster &95868A908 &\$13.79 &1 &\$13.79 \\
Nylon Screw (1/4-20) &Fasten Relay Circuit &McMaster &95868A744 &\$15.78 &1 &\$15.78 \\
Nylon Nut (1/4-20) &Fasten Relay Circuit &McMaster &94812A700 &\$13.24 &1 &\$13.24 \\
Nylon Nut (6-40) &Fasten Trap to Enclosure &McMaster &94812A663 &\$12.95 &1 &\$12.95 \\ \hline \hline
\multicolumn{6}{|r|}{\textbf{Total Table Price:}} &\textbf{\$5,704.41} \\
\hline
\bottomrule
\end{tabular}
\end{table*}

 \begin{table*}[!htp]\centering
\caption{Circuit Parts}\label{tab: circuit}
\scriptsize
\begin{tabular}{||l r r r r r r|}
\hline
\textbf{Item} & \textbf{Application} &\textbf{Supplier} &\textbf{Part Number} &\textbf{Price/Item} & \textbf{Qty.} & \textbf{Price} \\ \hline\hline
\multicolumn{7}{|c|}{\textbf{Miscellaneous Parts}}
\\ \midrule \hline
Ring Terminals &Attach Wires to Board &McMaster &7113K552 &\$18.68 &1 &\$18.68 \\
Sconnector kitle &Attach HV-DC to Circuit &Digikey &3985-PE33018-12-ND &\$109.99 &1 &\$109.99 \\
HV Wire (10 ft) &Connect Components &McMaster &8296K11 &\$54.80 &1 &\$54.80 \\
SHV Terminal &Attach HV-DC to Circuit &Digikey &A24669-ND &\$26.33 &2 &\$52.66 \\
5 V DC &Power Relay Circuit &Amazon &B09NLMVXMZ &\$7.98 &1 &\$7.98 \\
Banana Plug M (W) &Wires to Banana Plugs &Digikey &J144-ND &\$2.65 &2 &\$5.30 \\
Banana Plug M (R) &Wires to Banana Plugs &Digikey &J145-ND &\$2.74 &2 &\$5.48 \\
Banana Plug M (G) &Wires to Banana Plugs &Digikey &J340-ND &\$2.83 &2 &\$5.66 \\
Banana Plug M (BK) &Wires to Banana Plugs &Digikey &J146-ND &\$2.71 &2 &\$5.42 \\
Banana Plug M (B) &Wires to Banana Plugs &Digikey &J344-ND &\$1.52 &2 &\$3.04 \\
Banana Plug M (O) &Wires to Banana Plugs &Digikey &J341-ND &\$2.58 &2 &\$5.16 \\
Banana Plug M (GY) &Wires to Banana Plugs &Digikey &J346-ND &\$2.29 &2 &\$4.58 \\
Banana Plug M (PL) &Wires to Banana Plugs &Digikey &J345-ND &\$2.33 &2 &\$4.66 \\
Banana Plug M (Y) &Wires to Banana Plugs &Digikey &J342-ND &\$2.72 &2 &\$5.44 \\
Banana Plug F (W) &Wires to Banana Plugs &Digikey &J150-ND &\$1.38 &2 &\$2.76 \\
Banana Plug F (R) &Wires to Banana Plugs &Digikey &J151-ND &\$1.31 &2 &\$2.62 \\
Banana Plug F (G) &Wires to Banana Plugs &Digikey &J153-ND &\$1.45 &2 &\$2.90 \\
Banana Plug F (BK) &Wires to Banana Plugs &Digikey &J118-ND &\$1.31 &2 &\$2.62 \\
Banana Plug F (B) &Wires to Banana Plugs &Digikey &J155-ND &\$1.44 &2 &\$2.88 \\
Banana Plug F (O) &Wires to Banana Plugs &Digikey &J356-ND &\$1.20 &2 &\$2.40 \\
Banana Plug F (GY) &Wires to Banana Plugs &Digikey &J359-ND &\$1.28 &2 &\$2.56 \\
Banana Plug F (PL) &Wires to Banana Plugs &Digikey &J358-ND &\$1.33 &2 &\$2.66 \\
Banana Plug F (Y) &Wires to Banana Plugs &Digikey &J154-ND &\$1.11 &2 &\$2.22 \\
10 M Resistors &Current Limiter &Digikey &4506-HVLR3908F10M0K9-ND &\$14.03 &2 &\$28.06 \\
1 M Resistors &Current Limiter &Digkiey &VR68J1.0MCT-ND &\$0.93 &2 & \$1.86 \\
HV Tape &Apply on Solder Joints &McMaster &7682A42 &\$31.07 &1 &\$31.07 \\
Shrink Tubing Kit &Cover Solder Joints &Digikey &298-11579-ND &\$99.25 &1 &\$99.25 \\
\hline \multicolumn{7}{|c|}{\textbf{Planar Electrode Circuit}} \\ \hline
Test Point &Use to Probe Electrodes &Digikey &36-5199CT-ND &\$0.33 &1 &\$0.33 \\
Black Terminal &Wires to Electrodes &Digikey &1849-1168-1-ND &\$1.26 &3 &\$3.78 \\
Red Terminal &Wires to Electrodes &Digikey &1849-1170-1-ND &\$1.26 &1 &\$1.26 \\
Blue Terminal &Wires to Electrodes &Digikey &1849-1156-1-ND &\$1.26 &1 &\$1.26 \\
Green Terminal &Wires to Electrodes &Digikey &1849-1155-1-ND &\$1.26 &1 &\$1.26 \\
White Terminal &Wires to electrodes &Digikey &1849-1171-1-ND &\$1.26 &1 &\$1.26 \\
Yellow Terminal &Wires to Electrodes &Digikey &1849-1169-1-ND &\$1.26 &1 &\$1.26 \\
Planar Trap Board PCB &Ion Trapping &OSH Park &N.A. &\$77.30 &1 &\$77.30 \\
SHV to BNC Adapter &Ground SHV &ThorLabs &T4004 & \$50.5 & 1 & \$50.50 \\ 
BNC Terminator &Ground SHV &L-com &BM50G-1W &\$11.59 &2 & \$23.18 \\
\hline 
\multicolumn{7}{|c|}{\textbf{Relay Circuit}} \\ \hline
Relay Circuit PCB &Give Structure to Circuit &Osh Park &N.A. &\$285.00 &1 &\$285.00 \\
PCB Stencil &Help with Soldering &Osh Stencils &N.A. &\$29.09 &1 &\$29.09 \\
Solder Paste &Solder SMD Components &Digikey &315-NC191LT10-ND &\$7.95 &1 &\$7.95 \\
Reed Relays &Shuttling Voltages &Digikey &306-1193-ND &\$12.05 &10 &\$120.50 \\
Terminal Block (7) &Output Interface &Digikey &A135925-ND &\$4.48 &1 &\$4.48 \\
Terminal Block (3) &Input Interface &Digikey &A98085-ND &\$1.70 &1 &\$1.70 \\
1 M Resistor &N.A. &Digikey &A138410CT-ND &\$0.22 &16 &\$3.57 \\
100 M Resistor &N.A. &Digikey &A105971CT-ND &\$0.43 &6 &\$2.58 \\
Diodes &Flyback Diodes for Relays &Digikey &ZLLS410CT-ND &\$0.37 &10 &\$3.67 \\
5 V Jack &Power MOSFETS &Digikey &2073-DCJ250-05-A-K1-ACT-ND &\$1.02 &1 &\$1.02 \\
300 $\Omega$ Resistor &N.A. &Digikey &A121216CT-ND &\$0.90 &10 &\$9.01 \\
3.3 k Resistor &N.A. &Digikey &408-1639-1-ND &\$2.73 &10 &\$27.26 \\
MOSFET &Switch Reed Relays &Digikey &NTA7002NT1GOSCT-ND &\$0.16 &10 &\$1.61 \\
Signal Terminal (12) &Signal Transfer &Digikey &2057-SMC-1-12-1-GT-ND &\$0.90 &1 &\$0.90 \\
\hline\hline 
\multicolumn{6}{|r|}{\textbf{Total Table Price:}} & \textbf{\$1,113.76} \\ 
\hline
\bottomrule
\end{tabular}
\end{table*}

\FloatBarrier
\clearpage
\end{singlespace}

\end{document}